\documentclass[10pt,conference]{IEEEtran}

\usepackage{iftex}
\ifXeTeX
  \usepackage{fontspec}
\fi

\usepackage{amsmath,amssymb,amsfonts}
\usepackage{booktabs}
\usepackage{cite}
\usepackage{graphicx}
\usepackage{url}
\usepackage{xcolor}
\usepackage[most]{tcolorbox}
\usepackage{algorithm}
\usepackage{algpseudocode}
\usepackage{multirow}
\usepackage[most]{tcolorbox}
\usepackage{fontawesome5}
\usepackage{hyperref}
\usepackage[caption=false,font=footnotesize]{subfig}

\definecolor{sagegreen}{RGB}{138,179,137}
\colorlet{boxcolor}{sagegreen}

\newif\ifshowtodos
\showtodostrue

\newcommand{\methodname}{\textsc{SpecBuild}}

\newtcolorbox{findingbox}[1][]{
  enhanced,
  attach boxed title to top left={xshift=4mm,yshift=-2mm},
  colback=boxcolor!10,
  colframe=boxcolor!60,
  colbacktitle=boxcolor!80,
  coltitle=white,
  fonttitle=\bfseries\small,
  boxed title style={size=small, colframe=boxcolor!80, sharp corners},
  sharp corners,
  boxrule=0.8pt,
  left=4pt, right=4pt, top=4pt, bottom=4pt,
  breakable,
  title={#1}
}

\begin{document}

\title{Speculative Forking: Turning Idle Wait into Parallel Exploration for Repository Environment Setup}

\author{
\IEEEauthorblockN{
Yuling Shi\textsuperscript{1},
Mingchu Zhong\textsuperscript{1},
Zhensu Sun\textsuperscript{2},
Weihan Peng\textsuperscript{1},
Zhi Chen\textsuperscript{2},
Shilin He\textsuperscript{3},\\
Chengcheng Wan\textsuperscript{4}\textsuperscript{5},
Kai Cai\textsuperscript{6},
Hongyu Zhang\textsuperscript{7},
David Lo\textsuperscript{2},
Xiaodong Gu\textsuperscript{1}}
\IEEEauthorblockA{
\textsuperscript{1}Shanghai Jiao Tong University\quad
\textsuperscript{2}Singapore Management University\\
\textsuperscript{3}The Chinese University of Hong Kong\\
\textsuperscript{4}East China Normal University\quad
\textsuperscript{5}Shanghai Innovation Institute\\
\textsuperscript{6}Independent Researcher\quad
\textsuperscript{7}Chongqing University\\
\texttt{\{yuling.shi,peng-weihan,xiaodong.gu\}@sjtu.edu.cn}\\
\texttt{\{zssun,zhi.chen.2023,davidlo\}@smu.edu.sg}\\
\texttt{\{selin.hsl,hopcookie,hongyujohn\}@gmail.com},
\texttt{ccwan@sei.ecnu.edu.cn}}
}

\maketitle
% \outline{Alternative title: ``Hiding the Latency of LLM-Based Environment Setup via Speculative Forking''.}
% \outline{Method-name note: final name TBD: SpecForge / SWE-Fork / SpecBuild.}

\begin{abstract}
Before developers can use a software project, they must first get it to run.
This often requires installing the right language runtime, system packages, build tools, and test commands.
Recent LLM-based agents automate environment setup by inspecting a repository, writing setup scripts, running them, and repairing the resulting errors.
However, these agents still follow a sequential trial-and-error loop: they try one setup route, wait for slow commands to finish, observe the failure, and only then decide what to try next.
Because environment commands often take long to run, this sequential workflow makes setup slow and limits exploration to one route at a time.

We present \methodname{}, an LLM-based environment setup system designed to be both faster and more successful through speculative forking.
\methodname{} explores candidate setup branches in parallel, returns the first committed environment, and discards the rest.
It applies this design at two granularities: strategy-level forking explores alternative setup routes, such as different base images and dependency policies, while command-level forking predicts failures of running commands and prepares repair branches before those failures are observed.

We evaluate \methodname{} on Multi-Docker-Eval, covering 334 tasks from 40 repositories and 9 programming languages.
Across three backbone models, \methodname{} improves the success rate by 6.4 to 12.0 percentage points over the strongest baseline, while reducing average setup time by 22.4\% to 41.6\%.
These results show that LLM-based environment setup need not trade speed for reliability: parallel speculative exploration can make agents both faster and more successful.
\end{abstract}

\section{Introduction}

\begin{figure*}[t]
\centering
\includegraphics[width=0.8\textwidth]{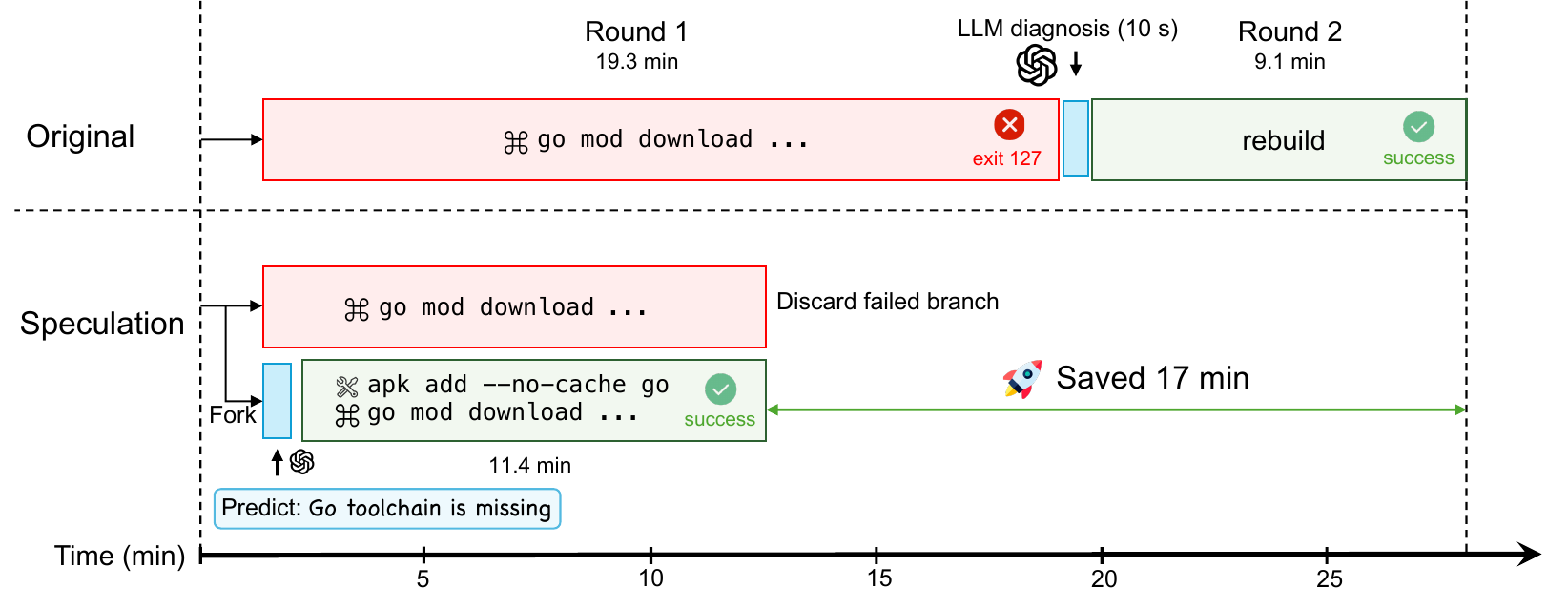}
\caption{Timeline comparison on \texttt{polarsignals/frostdb}.
The sequential agent waits for \texttt{go mod download} to fail before diagnosing and retrying, whereas speculation predicts the missing Go toolchain during the idle waiting interval and prepares an \texttt{apk}-based repair branch in advance.}
\label{fig:motivating-comparison}
\end{figure*}
Before developers can modify, test, debug, or reuse a software project, they must first get it to run.
Running a project's tests is central to validating code changes in both development and software-engineering research~\cite{ bouzenia2025you}.
This requires matching language runtimes, system packages, build tools, environment variables, and test commands to the project at hand.
Containerization makes this setup explicit through artifacts such as Dockerfiles, which have been widely studied as software-engineering and DevOps artifacts~\cite{henkel2020learning, cito2017empirical, eng2021revisiting}.
Yet writing these artifacts remains difficult: developers report Docker and Docker-Compose specifications as time-consuming, and prior work observes that they require system-administration knowledge and project-specific adaptation~\cite{reis2021developing, henkel2020learning, rosa2023automatically}.
Environment setup is therefore a practical bottleneck for developers before they can run tests, reproduce bugs, or validate fixes.

This bottleneck is amplified by recent efforts to make software-engineering tasks verifiable at scale.
Benchmarks and training pipelines for coding agents require executable environments in which tests can be run against real repositories~\cite{pan2024training, yang2026swe, jain2025r2e, yuan2026swe, chen2026securevibebench, shi2026swe}.
Recent systems now scale this construction to tens of thousands or even millions of environments~\cite{chen2026swe, fu2026davinci}.
In this setting, setup time is no longer a one-time inconvenience: each failed or slow construction attempt becomes a recurring wall-clock cost for evaluation, data generation, and developer workflows.

Recent work has started to automate environment setup with LLM-based agents.
These agents inspect repository files, generate setup scripts, run them in containers, read execution errors, and revise the scripts until the tests pass or a budget is exhausted~\cite{hu2026repo2run, bouzenia2025you, guo2025swe}.
Newer systems broaden this automation across languages, repositories, and training-data pipelines~\cite{li2026repolaunch, guo2026menvagent, guo2025swe}.
This feedback-driven loop also appears in broader software-engineering agents, which execute tools, observe feedback, and use that feedback to plan later actions~\cite{11029914, xia2024agentless, chen2026agentexecutor}.
In environment setup, however, the loop is typically sequential: the agent tries one setup route, waits for slow commands such as package installation, dependency resolution, and compilation to finish, observes the result, and only then decides what to try next~\cite{hu2026repo2run, bouzenia2025you, guo2025swe}.

This sequential structure exposes two uncertainties in environment setup.
First, the agent faces route uncertainty: it does not know which setup route, including choices such as the base image, runtime, package manager, and dependency policy, will build the repository.
A sequential loop can validate only one route at a time, so failed routes consume time and iteration budget before other plausible routes are tried.
Second, the agent faces command-outcome uncertainty: slow setup commands may fail, but a sequential loop waits for the failure before preparing a repair.
Together, these uncertainties couple wall-clock time and success: slow feedback makes each trial expensive, and one-route-at-a-time search limits how broadly the agent can explore.

Our insight is to address both uncertainties with \emph{speculative forking}.
Rather than waiting for one setup future to fail before considering another, \methodname{} explores candidate setup branches in parallel, returns the first committed environment, and discards the rest.
We apply this primitive at two granularities.
\emph{Strategy-level forking} explores alternative setup routes, such as different base images, runtimes, and dependency policies, addressing route uncertainty.
\emph{Command-level forking} predicts the failure of a currently running command and prepares a repaired branch before the failure is observed, addressing command-outcome uncertainty.
In both cases, speculation is predict-then-verify: branches only matter if real execution validates them.

We implement \methodname{} for LLM-based environment setup and evaluate it on Multi-Docker-Eval~\cite{fu-etal-2026-multi}, a benchmark of 334 environment-construction tasks drawn from 40 real-world repositories and 9 programming languages.
Across three backbone models, \methodname{} improves the success rate by $6.4$ to $12.0$ percentage points over the strongest baseline, while reducing average setup time by $22.4\%$ to $41.6\%$.
The results support the central claim of this paper: LLM-based environment setup need not trade speed for reliability, because parallel speculative exploration can make agents both faster and more successful.

In this paper, we make the following contributions:
\begin{itemize}
  \item \textbf{A dual objective for environment setup.}
  We formulate LLM-based environment setup as a dual-objective problem: improve
  the success rate while reducing wall-clock setup time. This framing
  captures the practical cost that developers and large-scale evaluation
  pipelines pay when sequential setup attempts fail slowly.

  \item \textbf{Speculative forking and its two granularities.}
  We introduce \emph{speculative forking}, a fork, explore, commit, and discard primitive over isolated setup branches, at two granularities: strategy-level forking races alternative setup routes, while command-level forking prepares repairs for likely command failures in advance.

  \item \textbf{Implementation and multi-language evaluation.}
  We implement \methodname{} and evaluate it on Multi-Docker-Eval across three
  backbone models and 9 programming languages, measuring end-to-end success and
  time, isolating each granularity's contribution, analyzing the branch-budget
  trade-off, and illustrating the mechanism through a case study. We release our
  implementation at \url{https://anonymous.4open.science/r/Speculative-Forking-C907}.
\end{itemize}

\begin{figure*}[t]
\centering
\includegraphics[width=\textwidth]{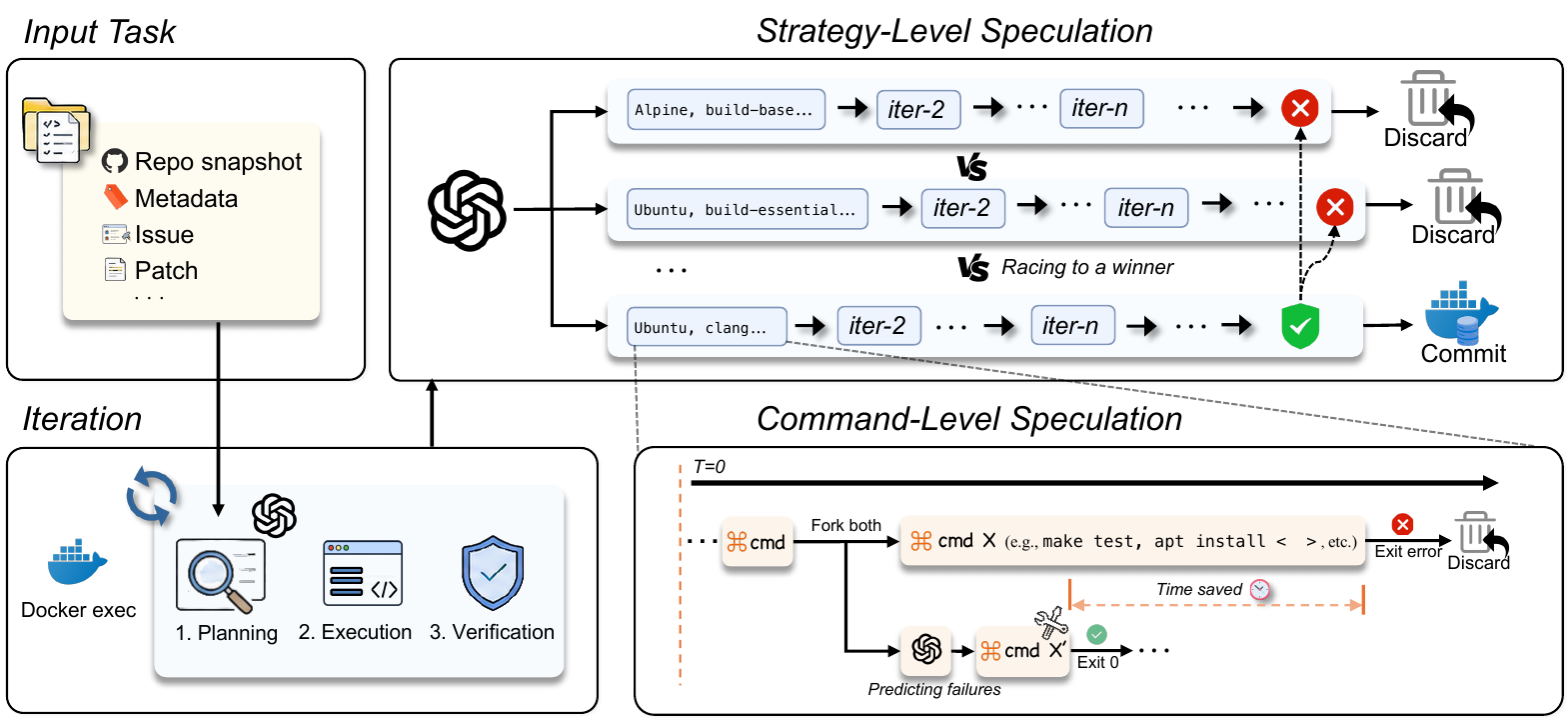}
\par\smallskip
{\footnotesize Strategy-level \textcolor{green!50!black}{$\checkmark$}: Agent considers setup complete.\par}
\caption{Overview of \methodname{}.
Given a repository, \methodname{} plans complementary setup routes that run as parallel branches, each repeatedly planning a setup script, executing it, and verifying the environment.
Inside execution, command-level speculation runs the original command and a candidate repair in parallel, adopting whichever validates.
The first committed environment is returned; the remaining branches are discarded.}
\label{fig:approach-overview}
\vspace{-0.5\baselineskip}
\end{figure*}

\section{Background and Motivation}
\label{sec:motivation}

Environment setup is a developer-facing problem before it is a benchmark abstraction.
Before developers can modify, test, or reproduce a project, they must first install the right language runtime, system packages, build tools, and commands.
Containerization makes this setup explicit through artifacts such as Dockerfiles, which are widely studied as software-engineering and DevOps artifacts~\cite{henkel2020learning, cito2017empirical, eng2021revisiting}.
Yet these artifacts remain difficult to write: developers report Docker and Docker-Compose specifications as time-consuming, while prior work observes that they often require system-administration knowledge and project-specific adaptation~\cite{reis2021developing, henkel2020learning, rosa2023automatically}.

\subsection{Problem Formulation}
\label{subsec:problem}

We study environment construction for verifiable software-engineering tasks~\cite{pan2024training,jain2025r2e}.
Following Multi-Docker-Eval~\cite{fu-etal-2026-multi}, each task provides a repository, a reported issue, and the developer's corresponding patch.
The system produces a setup configuration $(B,P,T)$: a base image $B$, installation commands $P$, and a test script $T$ for executing issue-related tests.

Environment correctness is assessed by whether the setup supports reproducing the issue and validating its fix.
We use the benchmark's fail-to-pass (F2P) criterion: the tests fail before the patch is applied and pass afterward under the same environment configuration.

Existing methods optimize their pipeline targeting the success rate alone.
Yet time matters in both day-to-day development and large-scale automation.
For a developer, each failed setup route blocks testing, reproduction, and validation until slow package installation, dependency resolution, or compilation commands finish~\cite{bouzenia2025you, rosa2023automatically}.
At scale, recent pipelines build executable environments to train and evaluate coding agents~\cite{pan2024training, yang2026swe, jain2025r2e, yuan2026swe}.
Recent environment-construction systems further scale this process to tens of thousands or millions of environments~\cite{chen2026swe, fu2026davinci}, where the per-instance build time becomes a first-order cost.
We therefore treat the wall-clock time to build the environment as a first-class objective, posing the task as two goals at once: raise the success rate while cutting build time.
To our knowledge, this is the first method to optimize the time cost of LLM-based environment construction rather than only report it.

\subsection{Motivating Example}
\label{subsec:motivating-example}

To see why this is hard, consider how current agents construct an environment.
A typical system such as SWE-Builder~\cite{guo2025swe} works in a loop: it inspects the repository's files (manifests, READMEs, CI configs), drafts a setup script, runs it in a container, reads whatever error the run produces, and revises the script, often re-inspecting the repository, before trying again.
It repeats this loop of analysis, execution, and diagnosis until the tests pass or a budget runs out.
The loop is sequential: the agent issues one script, waits for it to finish, and only then learns whether it worked.
This waiting is the structural bottleneck.
A single installation command may still take as long as it takes, but the sequential loop makes every next decision depend on that command's completion and failure.

On \texttt{polarsignals/frostdb}, a Go columnar database from Multi-Docker-Eval whose tests build only once a working Go toolchain is present, this sequential loop is slow, with progress blocked until each failure is observed (Figure~\ref{fig:motivating-comparison}).
Starting from a base image without the toolchain, the agent's first setup round ends in failure on the missing Go toolchain after $19.3$ minutes; it then reruns with Go installed, a further $9.1$ minutes, spending about $28.5$ minutes across two sequential rounds.
No useful progress is possible until that first failure surfaces, and no repair is ready when it does.

\methodname{} turns this waiting into useful work.
While the first round is still running, it predicts that the setup will fail for a missing Go toolchain and prepares the repair \texttt{apk add --no-cache go} on a parallel branch.
The prediction is platform-aware: seeing an Alpine base image, the model installs with \texttt{apk} rather than \texttt{apt}, and it does so from reasoning alone, with no cached history.
When the failure surfaces, the prepared branch already carries a working toolchain and is adopted directly, so the setup succeeds without a second round.
The task finishes in $11.4$ minutes, saving about $17$ minutes over the sequential run, and the committed environment passes the same F2P check.

This is speculation in the predict-then-verify sense: the predicted object is a command's outcome, and verification is its real execution.
At the higher level, \methodname{} also speculates over setup routes: it analyzes the repository, plans a primary route, and creates complementary routes that hedge against different environment assumptions.
The routes run independently, and the first committed environment is returned while the remaining branches are discarded.
Thus, strategy-level forking selects among plausible routes, while command-level forking hides wait-then-repair latency within a route.
We formalize speculative forking and its two granularities in Section~\ref{sec:approach}.

\section{Approach}
\label{sec:approach}

\subsection{Overview}
\label{subsec:overview}

\methodname{} replaces the sequential agent's one-route-at-a-time process with parallel speculative exploration, which at once hides the latency of slow commands and covers more of the configuration space than a single route.
% \todo{what is "hide"}
It analyzes the repository and plans several candidate routes (Section~\ref{subsec:strategy}), races them as parallel branches that each run an iterative setup loop, and within that loop speculates on the failure of individual commands (Section~\ref{subsec:outcome}); Section~\ref{subsec:guarantees} ties the two together in an end-to-end workflow.
We introduce \emph{speculative forking}, a design that explores setup alternatives before a sequential agent would reach them.

Figure~\ref{fig:approach-overview} overviews the design.
Given an input task, \methodname{} analyzes repository signals such as the snapshot, metadata, issue, and patch, and plans complementary setup routes.
At the strategy level, these routes run as parallel branches.
Each branch then follows the same iterative setup loop: plan a setup script, execute it in a container, and verify the resulting environment against the tests.
Inside the execution step, command-level speculation runs the original command and a candidate repair in parallel.
If the original command fails and the candidate repair succeeds, the repaired state is adopted immediately; otherwise the candidate repair is discarded.
The first committed environment is returned, and the non-winning branches are discarded.
This parallel breadth matches the structure of the problem: the configuration that builds is sparsely distributed across many routes, so exploring several at once reaches what a single sequential route cannot.

All exploration follows the same validation rule: run candidate work early, return the first committed environment, and discard branches that are no longer needed.
The two granularities differ only in what they fork and in what counts as reaching the goal.
Both are speculation in the predict-then-verify sense: work that may be needed is done ahead of time and verified by real execution, so a correct branch is kept and a wrong one is discarded.
The two granularities address two axes of uncertainty.
\emph{Strategy-level} forking (Section~\ref{subsec:strategy}) handles uncertainty \emph{between} routes, which configuration will build: it anticipates how a whole route could fail and hedges across routes accordingly.
\emph{Command-level} forking (Section~\ref{subsec:outcome}) handles uncertainty \emph{within} a route, whether a given command will fail, making an explicit per-command prediction of that failure.
Both anticipate failure and differ only in granularity, a whole route versus a single command.

\subsection{Strategy-Level Forking}
\label{subsec:strategy}

Strategy-level forking addresses uncertainty over which setup route will build the repository.
Real repositories rarely document their exact environment, so several plausible configurations coexist and which one builds is revealed only by trying.
A \emph{route} is a setup direction: a base-image family, a runtime version, a package manager, and a dependency policy.
For \texttt{frostdb}, three routes start from the official images \texttt{golang:1.20-alpine}, \texttt{golang:1.19-bookworm}, and \texttt{golang:1.19-bullseye}.
Rather than guess a single route, \methodname{} speculates across them: it pursues several plausible routes in parallel and lets execution decide which one builds.

\noindent\textbf{Generating routes.} An analysis phase reads the repository's build signals, such as manifests, CI test commands, and project metadata, and summarizes them into a structured profile.
Conditioned on this profile, \methodname{} plans a \emph{primary route}, the setup direction most likely to build the repository, and, by reasoning about how that route could fail, such as a wrong base-image family, an incompatible runtime version, or a mismatched package manager or dependency policy, composes a small set of \emph{complementary routes} that hedge those distinct failure modes.
Because the primary and its hedges are planned together before any route runs, they cover different environment assumptions by construction rather than by chance, which is what distinguishes route racing from drawing independent samples (Section~\ref{sec:rq2}).

\noindent\textbf{The branch iteration.} Each route runs as an isolated branch that repeats a three-step loop: it \emph{plans} a setup script from the profile, the route, and any feedback; \emph{executes} the script in a container; and \emph{verifies} the result using test commands or scripts selected or written by its agent.
A branch iterates this loop for up to $m$ rounds, and the diagnostics of a failed round condition the next.
It commits a candidate environment when its agent considers the setup complete; a branch that reaches the iteration limit without committing terminates without a submission.
Command-level speculation (Section~\ref{subsec:outcome}) operates inside the execute step of this same loop.

\noindent\textbf{Racing to a winner.} The branches run concurrently, bounded by a configurable budget of $k$ branches.
The first committed environment is returned, and the remaining branches are cancelled and discarded.
If no branch commits before all branches terminate or the task times out, no environment is submitted.
No route is privileged, so a failure on any single one never stalls the task.
This is where both gains originate: racing reaches a working configuration that a single route would miss, and it avoids the wall-clock a sequential agent spends exhausting dead ends one after another.
Because the branches race, the first to commit cancels the rest: no branch runs past the moment of commit, and a route that hits a dead end is abandoned within a few rounds.
The branch budget $k$ and the per-task speculation budget cap the total work, so racing explores breadth without unbounded cost; Section~\ref{sec:rq4} analyzes the effect of the branch budget.

\subsection{Command-Level Speculation}
\label{subsec:outcome}

Within the execute step of that same loop, individual commands run one after another.
Command-level forking addresses the uncertainty over whether a given command will fail, and turns the time spent running it, often minutes for an install or a compile, into preparation for its repair.
The design rests on an asymmetry: a wrong guess costs only a millisecond-scale fork and one short, bounded repair attempt, whereas a right guess removes an entire round of failure, diagnosis, and retry.
Because the downside is so much smaller than the upside, \methodname{} can speculate opportunistically.
On \texttt{frostdb}, the execute step runs the build, and \methodname{} anticipates that it will fail for a missing Go toolchain.

\noindent\textbf{Predicting failures.} Before a command runs, \methodname{} queries an LLM with the repository, its language, the base-image family, and the command, and receives a structured prediction: a failure signature and class, a candidate fix, and a confidence score.
For \texttt{frostdb} it predicts \texttt{GO\_MISSING} and proposes \texttt{apk add go}.
The choice is platform-aware: recognizing an Alpine base image, the model installs with \texttt{apk} rather than \texttt{apt}, and it does so from reasoning alone, with no cached history.

\noindent\textbf{When to speculate.} The value of speculation grows with a command's running time and its chance of failing, so \methodname{} speculates only on setup-relevant commands such as dependency installation, building, and compilation, and skips trivial ones like \texttt{cd} or \texttt{echo}.
A per-task budget bounds the number of predictions and their cost.

\noindent\textbf{Speculative repair via fork-both.} \methodname{} runs two attempts concurrently: one executes the original command, the other applies the predicted fix and then the command.
If the original succeeds, the repair is discarded; if the original fails while the repair succeeds, the repair is adopted, sparing a full diagnose-and-retry round; if both fail, the speculative repair is discarded and the branch continues from the original failure feedback.
A misprediction therefore merely wastes a bounded parallel attempt.
This is why fork-both is preferable to acting on a prediction directly: the predicted fix is verified against the real command rather than trusted.

\subsection{End-to-End Workflow and Validation}
\label{subsec:guarantees}

The two granularities run under a single driver.
Strategy-level forking launches the candidate routes as parallel branches and returns the first committed environment; each branch runs the plan, execute, and verify loop; and within a branch, risky commands may invoke command-level fork-both, which is the point where the two levels meet.

\iffalse
\begin{algorithm}[t]
\caption{\methodname{} Environment Construction}
\label{alg:specbuild}
\begin{algorithmic}[1]
\Require repository $repo$, branch budget $k$, iteration limit $m$
\Ensure a validated environment, or $\bot$
\State $profile \gets \Call{Analyze}{repo}$
\State $routes \gets \Call{Plan}{profile, k}$ \Comment{$k$ candidate routes (strategy-level)}
\State run \Call{RunBranch}{$r$} for all $r \in routes$ in parallel \label{line:race}
\State \textbf{commit} the first branch with $\mathrm{F2P}(env)$; discard the rest \label{line:commit}
%\Statex
\Function{RunBranch}{$route$}
    \For{$i \gets 1$ \textbf{to} $m$} \Comment{plan, execute, verify}
        \State $setup \gets \Call{Plan}{profile, route, feedback}$
        \For{\textbf{each} command $c$ in $setup$}
            \If{\Call{Risky}{$c$}}
                \State \Call{ForkBoth}{$c$} \Comment{command-level}
            \Else
                \State run $c$
            \EndIf
        \EndFor
        \If{$\mathrm{F2P}(env)$}
            \State \Return $env$
        \EndIf
        \State $feedback \gets$ diagnostics
    \EndFor
    \State \Return $\bot$
\EndFunction
\Statex
\Function{ForkBoth}{$c$}
    \State $fix \gets \Call{Predict}{c}$
    \State \textbf{in parallel:} run $c$ on fork $A$; run ($fix$, then $c$) on fork $B$
    \State \textbf{adopt} whichever succeeds; discard the other
\EndFunction
\end{algorithmic}
\end{algorithm}
\fi

Each branch follows its agent's own checks and completion criteria when deciding whether to commit.
Command-level forking preserves the original execution path while evaluating a repair in an isolated branch; a discarded repair therefore leaves the retained environment unchanged.
Speculation changes the timing and ordering of exploration while retaining the agent's checks and commit decision.
With a single branch and command-level speculation disabled, \methodname{} reduces to the sequential base agent.

\section{Experimental Setup}

We organize the evaluation around four research questions, followed by a case study.
\begin{itemize}
\item \textbf{RQ1 (Effectiveness).} How effectively does \methodname{} reduce the wall-clock time of environment construction while improving its success rate, compared with state-of-the-art agents?
\item \textbf{RQ2 (Ablation).} How much does each speculation level, strategy-level and command-level, contribute to the time and success gains?
\item \textbf{RQ3 (Generality).} How well do the gains generalize across the diverse programming-language ecosystems of the benchmark?
\item \textbf{RQ4 (Branch Sensitivity).} How many parallel branches does \methodname{} need?
\end{itemize}
The rest of this section describes the benchmark, baselines, metrics, and implementation used to answer them.

%\hy{if time allows, can analyze the Resource/computational cost somewhere}

\subsection{Benchmark}
\label{subsec:benchmark}

We evaluate all methods on Multi-Docker-Eval~\cite{fu-etal-2026-multi}, a benchmark designed specifically for automated environment construction.
It consists of 334 tasks drawn from 40 real-world repositories spanning 9 programming languages (C, C++, Go, Java, JavaScript, PHP, Python, Ruby, and Rust), each following the fail-to-pass criterion of Section~\ref{subsec:problem}.
This breadth of languages and build ecosystems within a single benchmark enables a controlled cross-language comparison under consistent evaluation conditions.

\subsection{Baselines}
\label{subsec:baselines}

We compare \methodname{} against four state-of-the-art environment-construction agents.
\textbf{Repo2Run}~\cite{hu2026repo2run} constructs executable Python environments from repository context; because it is Python-specific, we include it in the per-language analysis rather than the full multi-language comparison.
\textbf{RepoLaunch}~\cite{li2026repolaunch} automates dependency resolution, repository build, and test-result extraction across languages and platforms.
\textbf{SWE-Builder}~\cite{guo2025swe} inspects a repository, generates a setup script, runs it, and revises it from execution errors.
\textbf{MEnvAgent}~\cite{guo2026menvagent} retrieves environments from previously solved repositories, then reuses and patches them for a new project to avoid building from scratch.
RQ2 further includes two ablation controls, Sequential (More Budget) and Best-of-N (Naive Parallel), to test whether the gains come from a larger sequential budget or from running independent workers in parallel.

\begin{table*}[t]
  \centering
  \caption{End-to-end performance on the Multi-Docker-Eval benchmark.}
  \label{tab:rq1-main}
  \setlength{\tabcolsep}{6pt}
  \renewcommand{\arraystretch}{1.15}
  \begin{tabular}{llrrrrrr}
  \toprule
  Model & Method
    & \shortstack{Success\\Rate (\%)}
    & \shortstack{Commit\\Rate (\%)}
    & \shortstack{Avg.\\Time (min)}
    & \shortstack{Avg. Input\\Tokens (K)}
    & \shortstack{Avg. Output\\Tokens (K)}
    & \shortstack{Avg.\\Cost (\$)} \\
  \midrule
  \multirow{4}{*}{DeepSeek-v4-Flash}
    % & Repo2Run      & 3.59 & 9.88 & 53.60 & 181.87 & 4.75 & 0.0128 \\
    & RepoLaunch    & 11.08 & 42.81 & 36.87 & 796.89 & 8.87 & 0.0359 \\
    & SWE-Builder   & 17.96 & 34.13 & 66.75 & 142.59 & 35.54 & 0.0160 \\
    & MEnvAgent     & 24.47 & 45.41 & 48.01 & 235.23 & 114.28 & 0.0402 \\
    & \methodname{} & \textbf{30.84} & 46.71 & \textbf{28.04} & 775.37 & 78.06 & 0.0483 \\
  \midrule
  \multirow{4}{*}{MiMo-v2.5-Pro}
    % & Repo2Run      & 4.49 & 10.48 & 51.45 & 179.00 & 2.78 & 0.0064 \\
    & RepoLaunch    & 14.67 & 63.77 & 55.80 & 1313.27 & 15.84 & 0.1395 \\
    & SWE-Builder   & 21.56 & 51.20 & 59.98 & 195.37 & 53.02 & 0.0737 \\
    & MEnvAgent     & 26.65 & 61.38 & 49.38 & 322.31 & 170.49 & 0.1819 \\
    & \methodname{} & \textbf{38.02} & 56.89 & \textbf{33.58} & 832.26 & 64.45 & 0.2059 \\
  \midrule
  \multirow{4}{*}{GPT-5.5}
    % & Repo2Run      & TBD & TBD & TBD & TBD & TBD & TBD \\
    & RepoLaunch    & 23.35 & 62.27 & 46.06 & 884.43 & 9.65 & 4.2670 \\
    & SWE-Builder   & 32.33 & 60.77 & 64.88 & 113.87 & 29.92 & 1.4434 \\
    & MEnvAgent     & 46.40 & 61.97 & 26.78 & 88.74 & 18.34 & 0.9780 \\
    & \methodname{} & \textbf{58.38} & 74.85 & \textbf{20.79} & 254.79 & 66.79 & 3.1168 \\
  \bottomrule
  \end{tabular}
  \end{table*}
\label{sec:rq1}

\subsection{Metrics}
\label{subsec:metrics}

Following Multi-Docker-Eval~\cite{fu-etal-2026-multi}, we report five metrics for each method.
\begin{itemize}
\item \textbf{Success Rate:} the percentage of tasks whose constructed environment is verified as correct. Following Multi-Docker-Eval~\cite{fu-etal-2026-multi}, verification assesses whether the environment can execute the issue-related tests to reproduce the reported issue and validate its fix, as described in Section~\ref{subsec:problem}.

\item \textbf{Commit Rate:} the percentage of tasks for which the agent commits an environment for evaluation, a self-reported signal that the agent judged its setup ready; it does not require the environment to verify the change, so a high Commit Rate alone does not imply success.
\item \textbf{Average Time:} the average wall-clock time per task; our primary efficiency metric.
\item \textbf{Token Consumption:} the average input and output tokens used per task during environment construction.
\item \textbf{Average Cost:} the average LLM API cost per task in US dollars.
\end{itemize}

\subsection{Models and Implementation}
\label{subsec:implementation}

To assess robustness across reasoning backbones, we run \methodname{} and all baselines end-to-end with three state-of-the-art LLMs spanning open-source and proprietary models: DeepSeek-V4-Flash~\cite{deepseekai2026deepseekv4}, MiMo-V2.5-Pro~\cite{mimo2026v25pro}, and GPT-5.5~\cite{openai2026gpt55}.
Unless noted otherwise, \methodname{} uses a branch budget of $k=5$, an iteration limit of $m=15$ per branch, a per-task budget of $15$ speculative predictions, temperature $0$, and a $5400$\,s timeout.
All experiments run on a single server with two Intel Xeon Platinum 8259CL CPUs ($96$ threads) and $128$\,GiB of RAM under Ubuntu 22.04.

\section{Experimental Results}
\label{sec:results}
\subsection{RQ1: Effectiveness}

\noindent\textbf{Setup.} We compare \methodname{} against all baselines on the full Multi-Docker-Eval set of 334 tasks across the three backbone LLMs, reporting Success Rate, Commit Rate, and average time per task (Table~\ref{tab:rq1-main}), under the configuration of Section~\ref{subsec:implementation}.

% \hy{should be "Experimental Results" here? after Setup. Same for other RQs}
\noindent\textbf{\methodname{} is both faster and more successful.} Across all three backbones, \methodname{} attains the highest Success Rate and the lowest average time of any method. Against the strongest baseline, MEnvAgent in every case, it raises the Success Rate from $24.5\%$ to $30.8\%$ on DeepSeek-v4-Flash, from $26.7\%$ to $38.0\%$ on MiMo-v2.5-Pro, and from $46.4\%$ to $58.4\%$ on GPT-5.5, a gain of $6.4$ to $12.0$ percentage points. At the same time it cuts average setup time from $48.0$ to $28.0$, from $49.4$ to $33.6$, and from $26.8$ to $20.8$ minutes respectively, a reduction of $22.4\%$ to $41.6\%$. Both gains are statistically significant. On every backbone, the Success Rate improvement over the strongest baseline is significant under McNemar's test~\cite{mcnemar1947note}, and the setup-time reduction under the Wilcoxon signed-rank test~\cite{wilcoxon1945individual}, each paired per task at $p<0.005$.

\noindent\textbf{The two goals do not trade off.} No baseline is at once the fastest and the most successful: SWE-Builder and RepoLaunch are both slow and far less accurate, while MEnvAgent, the most successful baseline, remains much slower than \methodname{}. \methodname{} is the only method that leads on both axes, the direct payoff of replacing sequential exploration with parallel speculative exploration.

\noindent\textbf{Commit Rate reflects submission, not success.} Commit Rate counts whether the agent submitted an environment it judged ready, which need not coincide with the environment actually working: on MiMo-v2.5-Pro, RepoLaunch commits more often ($63.8\%$) than \methodname{} ($56.9\%$) but verifies far fewer tasks ($14.7\%$ versus $38.0\%$ Success Rate). We therefore report Commit Rate for completeness and read the Success Rate as the measure of success (Section~\ref{sec:rq3}).

\begin{findingbox}[Finding 1]
Across three backbones, \methodname{} raises the Success Rate by 6.4 to 12.0 percentage points over the strongest baseline while cutting average setup time by 22.4\% to 41.6\%, the only method that is at once the fastest and the most successful.
\end{findingbox}

\subsection{RQ2: Ablation}
\label{sec:rq2}

\noindent\textbf{Setup.} We decompose \methodname{}'s gain on Multi-Docker-Eval.
Building up from a sequential agent, we first add strategy-level forking and then add command-level forking on top, so the third row is the full \methodname{} configuration.
We also add two controls that rule out alternative explanations of the gain:
\begin{itemize}
\item \textbf{Sequential (More Budget):} the sequential agent with a larger budget, set to a comparable token-cost budget, ruling out that the gain comes merely from spending more.
\item \textbf{Best-of-N (Naive Parallel):} five independent base-agent workers run in parallel, matching \methodname{}'s maximum strategy budget, ruling out that the gain comes merely from running more workers.
\end{itemize}
% \hy{yes I think this experiment is important, to show that it isn't  simply running multiple candidates simultaneously. Also, in the paper you can highlight it too. Otherwise, people may think it is a bit too simple.}

\begin{table}[t]
\centering
\caption{Component ablation and controls (DeepSeek-v4-Flash).}
\label{tab:ablation}
\resizebox{\columnwidth}{!}{%
\begin{tabular}{lcccc}
\toprule
Configuration & Success Rate & Commit & Time (min) & Cost (\$) \\
\midrule
\multicolumn{5}{@{}l}{\textit{Component ablation}} \\
Sequential                  & 22.2\% & 31.4\% & 57.4 & 0.0144 \\
\quad + Strategy             & 29.9\% & 42.2\% & 32.7 & 0.0449 \\
\quad + Strategy + Command   & \textbf{30.8\%} & \textbf{46.7\%} & \textbf{28.0} & 0.0483 \\
\midrule
\multicolumn{5}{@{}l}{\textit{Controls}} \\
Sequential (More Budget)    & 25.1\% & 35.9\% & 81.3 & 0.0501 \\
Best-of-N (Naive Parallel)  & 24.8\% & 35.3\% & 52.6 & 0.0582 \\
\bottomrule
\end{tabular}%
}
\end{table}

\begin{figure*}[t]
  \centering
  \subfloat[DeepSeek-V4-Flash\label{fig:cross-language-deepseek}]{%
    \includegraphics[width=0.315\textwidth]{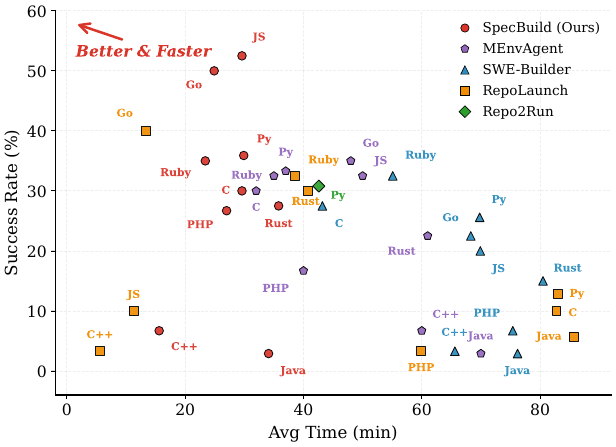}%
  }\hfill
  \subfloat[Mimo-V2.5-Pro\label{fig:cross-language-mimo}]{%
    \includegraphics[width=0.315\textwidth]{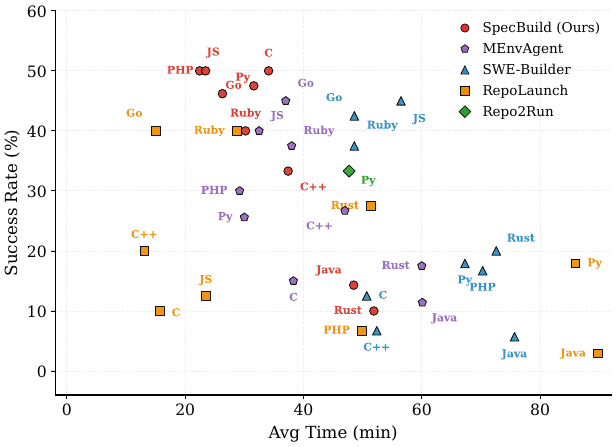}%
  }\hfill
  \subfloat[GPT-5.5\label{fig:cross-language-gpt}]{%
    \includegraphics[width=0.315\textwidth]{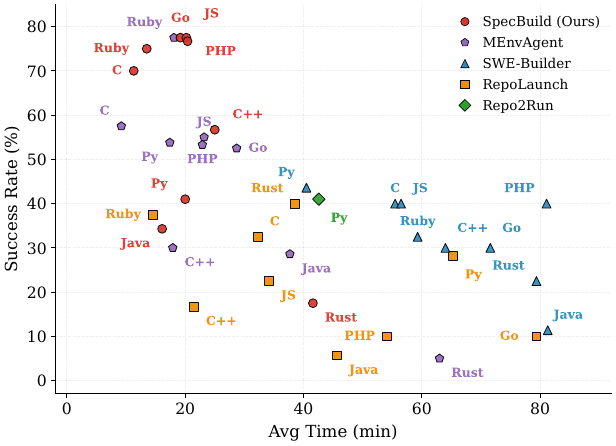}%
  }
  \caption{Cross-language comparison on Multi-Docker-Eval: each point is a (method, language) pair; upper-left (higher Success Rate, lower time) is better.}
  \label{fig:cross-language}
  \end{figure*}

\noindent\textbf{The two levels play complementary roles.}
Adding strategy-level forking primarily improves success, raising the Success Rate from $22.2\%$ to $29.9\%$, while also cutting average time from $57.4$ to $32.7$ minutes.
This matches its purpose: strategy-level forking addresses route uncertainty by validating several problem-guided setup routes instead of spending the whole budget on one route.
Adding command-level forking on top gives the full system, which further cuts average time to $28.0$ minutes and raises Success Rate to $30.8\%$.
This matches its purpose as well: command-level forking addresses command-outcome uncertainty by preparing repairs before slow commands fail.

\noindent\textbf{More sequential budget does not replace route exploration.}
Could the gain be explained by \methodname{} simply spending more?
The larger-budget sequential control reaches only $25.1\%$ Success Rate, $5.7$ pp below \methodname{}, while taking $81.3$ minutes and costing about the same as the full system.
This shows that deeper sequential search does not resolve route uncertainty: it spends more effort on one trajectory, whereas \methodname{} uses the budget to validate multiple setup routes.

\noindent\textbf{Naive parallelism is not enough.}
Could the gain be explained by parallelism alone?
We build a Best-of-N control where each task launches five independent base-agent workers, matching \methodname{}'s maximum strategy budget.
As in \methodname{}, the first worker to commit a candidate environment determines the output.
Best-of-N improves over the sequential agent, raising Success Rate from $22.2\%$ to $24.8\%$ and Commit Rate from $31.4\%$ to $35.3\%$, showing that parallel sampling increases candidate coverage.
However, it remains well below \methodname{}'s $30.8\%$ Success Rate and $46.7\%$ Commit Rate, while taking $52.6$ minutes per task and costing $\$0.0582$, compared with \methodname{}'s $28.0$ minutes and $\$0.0483$.
Thus, the gain comes from problem-guided speculation, which allocates complementary setup routes and prepares likely repairs, not from launching independent workers alone.

\begin{findingbox}[Finding 2]
Strategy-level forking mainly improves success, command-level forking further reduces latency, and neither comparable sequential budget nor 5-way naive parallelism matches the full system.
\end{findingbox}

\subsection{RQ3: Generality}
\label{sec:rq3}

\noindent\textbf{Setup.} We test whether \methodname{}'s gains hold across language ecosystems rather than only within one.
We disaggregate the results by language and report the Success Rate, Commit Rate, and average time of each (method, language) pair (Figure~\ref{fig:cross-language}).
%\todo{the ``two panels'' phrasing should then map to the Success-Rate and Commit-Rate columns.}

\noindent\textbf{The gains generalize across languages.} \methodname{}'s points cluster in the upper-left region across the language spectrum, while the baselines spread toward the lower-right: it leads on Success Rate in most languages and completes them in $23$ to $36$ minutes, against the $43$ to $81$ minutes of SWE-Builder.
The gains hold across ecosystems rather than reflecting tuning to any single one.

\noindent\textbf{Success Rate, not Commit Rate, separates the methods.} The template-based RepoLaunch and the Python-specialized Repo2Run reach high Commit Rate on the languages they cover, yet their Success Rate stays low: they produce a runnable container that is mistaken for readiness while the test suite remains non-executable.
\methodname{} ranks high on both, producing environments that not only run but pass the fail-to-pass check that downstream use requires.

\noindent\textbf{Where the gain narrows.} The benefit is smaller on languages dominated by system-level configuration.
Go, Python, and JavaScript, with declarative ecosystems and mature package managers, admit a small space of plausible setups that parallel exploration covers efficiently; C++, Java, and Rust instead hinge on system libraries, compiler toolchains, and version-specific builds that are hard to infer, where all methods score low.
The contrast is sharp even within one method: RepoLaunch attains $40.0\%$ Success Rate on Go in $13.4$ minutes, where a unified toolchain (\texttt{go build}, \texttt{go mod}) maps to a deterministic build, but drops to $10.0\%$ on C while taking $82.8$ minutes, over six times longer, as fragmented build systems (CMake, Make, Autotools) force an open-ended search.
Here the bottleneck is resolving underspecified build requirements, a challenge orthogonal to search breadth that parallel exploration complements rather than supplants; it narrows but does not reverse \methodname{}'s advantage.

\begin{findingbox}[Finding 3]
\methodname{}'s success and time gains hold across all nine ecosystems of Multi-Docker-Eval, narrowing only where the bottleneck shifts from search breadth to underspecified system-level builds (C++, Java, Rust).
\end{findingbox}

% (per-language results table: a student will add the multilingual version; see the \todo in the RQ3 Setup)

\subsection{RQ4: Branch Sensitivity}
\label{sec:rq4}

\noindent\textbf{Setup.} By default, \methodname{} does not fix the number of branches: its planner proposes between two and five candidate routes per repository and adapts the count to the project. To study how this budget shapes the cost-benefit trade-off, we sweep a fixed $K$ from $1$ to $5$ under DeepSeek-V4-Flash.

\begin{figure}[t]
\centering
\includegraphics[width=0.75\columnwidth]{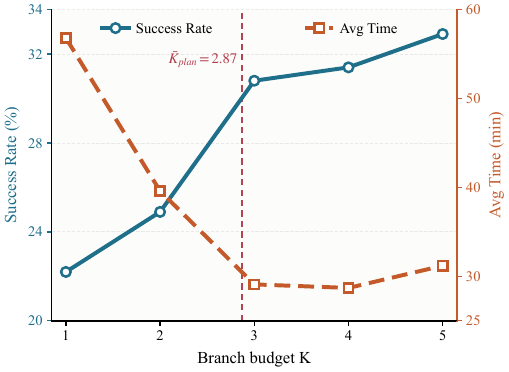}
\vspace{-0.2cm}
\caption{Sensitivity results under DeepSeek-V4-Flash for $K=1$ to $5$.}
\vspace{-0.2cm}
\label{fig:k-sensitivity}
\end{figure}

\noindent\textbf{The planner self-selects a cost-effective budget.} Sweeping fixed $K$ (Figure~\ref{fig:k-sensitivity}), Success Rate climbs steeply from $22.2\%$ at $K=1$ to $30.8\%$ at $K=3$ and then rises only gradually, while average setup time falls sharply from $56.8$ to about $29$ minutes by $K=3$ and stays near its minimum.
Because each branch runs its own planning and repair loop, token cost grows with $K$, so the marginal success beyond $K\approx3$ is bought at a disproportionate cost.
The knee is therefore around $K=3$, where most of the success gain is realized at near-minimum time while cost stays moderate.
Average time edges up slightly at $K=5$ largely because the extra environments solved only at a larger budget are the harder, slower ones that smaller budgets abandon, which lifts the mean over solved tasks.
\methodname{}'s adaptive planner averages $\bar{K}_{plan}=2.87$ routes, about three, landing right at this knee, so it captures most of the multi-route benefit without a tuned budget.

\begin{findingbox}[Finding 4]
\methodname{}'s planner automatically lands at the efficiency knee, averaging about three routes where fixed-$K$ success gains flatten and setup time is near its minimum, so it captures most of the multi-route benefit without tuning.
\end{findingbox}

\begin{figure*}[t]
  \centering
  \includegraphics[width=0.8\linewidth]{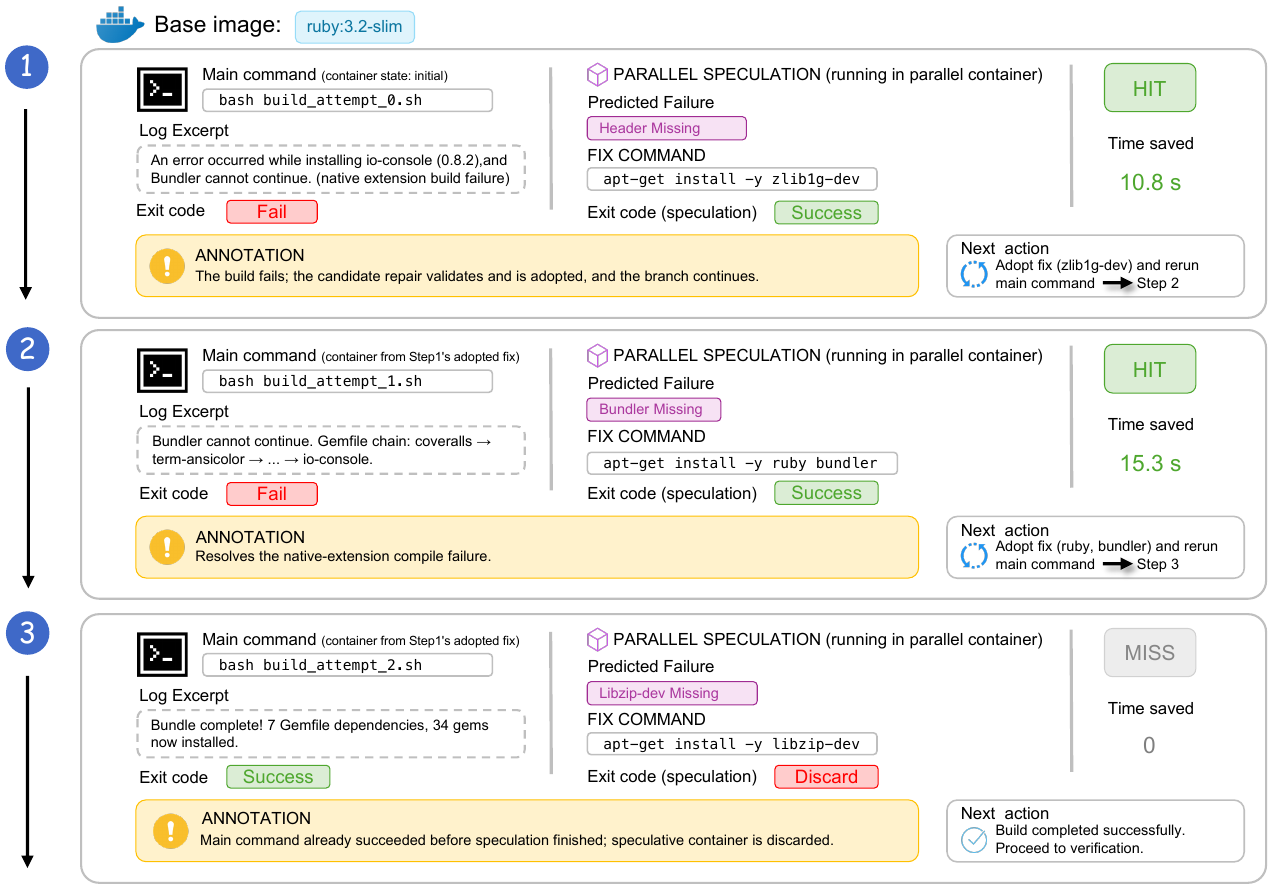}
  \vspace{-0.25cm}
  \caption{Case study: three command-level speculations on one \texttt{rubyzip} branch, two adopted and one discarded.}
  % \caption{Case study on a \texttt{rubyzip} branch: at each risky build command, \methodname{} runs the original command and a candidate repair in parallel and adopts whichever the build validates.}
  \label{fig:rq4-case}
  \vspace{-0.25cm}
\end{figure*}

\subsection{Case Study}
\label{sec:casestudy}

\noindent This case study illustrates the command-level mechanism rather than the branch budget: we trace one branch to show how a prepared repair is validated by real execution and then adopted or discarded.

\noindent\textbf{A hit is an adopted branch, not a correct diagnosis.}
Figure~\ref{fig:rq4-case} traces one \texttt{rubyzip} (Ruby) branch on the \texttt{ruby:3.2-slim} image, where \texttt{bundle install} fails while compiling a native gem extension.
At each risky build, \methodname{} runs the original command and a candidate repair, a system-package install such as \texttt{apt-get install zlib1g-dev}, side by side and adopts whichever the build validates.
The guess need not name the true cause: at each step the candidate repair runs in parallel with the original command, and the branch adopts whichever validates, converging to a working environment without a serial diagnose-and-retry detour; the third speculation is unnecessary, because the build succeeds on its own, and is discarded for the cost of a fork.
Because adoption is decided by real execution rather than by the prediction, a wrong or unnecessary bet is discarded without ever changing the environment that is finally committed.
This illustrates how an adopted repair removes a retry round while a discarded repair leaves the retained environment unchanged.

\noindent\textbf{Headroom that grows with model capability.} Because command-level speculation removes a round only when a prepared repair validates a failing command, the latency it hides grows with predictor quality.
More accurate predictions create more opportunities to prepare successful repairs before the original commands fail.
Ecosystems with canonical fixes, such as Go, already show what a strong predictor delivers, while more open-ended ones, such as Python, leave headroom that more capable backbones close.
Command-level speculation therefore converts advances in model reasoning directly into faster setup, and stands to gain as backbone models improve.
A perfect predictor would catch every recoverable failure and remove every corresponding round, the ceiling this mechanism approaches as models improve.
% \todo{Compute the oracle upper bound (latency removed if every recoverable failure were predicted and repaired) and report the gap to the current predictor.}

\section{Discussion}

\label{sec:discussion}

\subsection{Command-Level Speculation as a Design Space}

Command-level speculation is best seen not as a single heuristic but as a cheap, discardable bet with an asymmetric payoff: a prepared repair is adopted when real execution validates it and discarded otherwise, so correctness never depends on the guess.
This opens a concrete design space for future improvement: better failure predictors, stronger repair generators, and policies trained on setup traces would each raise how often a bet is adopted and so hide more latency.
Speculative forking thus extends a long line of predict-then-verify systems, from speculative execution to recent speculative LLM and agent-serving work~\cite{10.1145/1411204.1411205, leviathan2023fast,ye2025speculative}.

\subsection{When Speculation Pays Off}

Speculation is most valuable when the setup process exposes long, uncertain commands with structured failure modes.
Package installation, dependency resolution, compilation, and test preparation often take minutes before revealing a missing toolchain, a stale module graph, or an absent system library, recurring bottlenecks in environment construction and test-execution agents~\cite{bouzenia2025you,fu-etal-2026-multi, rosa2023automatically}.
In such cases, a correct speculative branch can remove a full wait, diagnose, and retry round from the critical path.
The largest gains therefore arise when commands are long enough to expose useful waiting time and when failure modes are regular enough for advance repair work to be validated.
This gives a practical scheduling rule: speculate where there is enough latency and enough regularity to make advance work worth doing.
This rule is also a deployment knob: where wall-clock latency dominates, the branch and speculation budgets control how much parallel exploration to spend, and \methodname{} sets this adaptively rather than by hand (Section~\ref{sec:rq4}).

\subsection{Route Racing versus Naive Parallelism}

Strategy-level forking may look like simply running several setup attempts in parallel, but the advantage is not parallelism alone.
Each branch is a distinct, problem-guided setup route rather than an interchangeable sample, and our ablation isolates the difference: at matched cost, route racing outperforms both a larger sequential budget and naive parallelism (Section~\ref{sec:rq2}).
Where a sequential agent deepens a single setup attempt from execution feedback~\cite{hu2026repo2run,bouzenia2025you,guo2025swe}, route racing selects the first route whose agent commits.

Command-level forking addresses a different source of latency.
It does not race complete routes; it predicts that a running command is likely to fail and prepares a repair before the failure is observed.
Together, the two granularities change the schedule of execution-driven setup: strategy-level forking broadens the search over routes, while command-level forking turns command waiting time into repair preparation.
Both remain bounded by branch and speculation budgets, and both commit only work validated by real execution.

\section{Threats to Validity}

% \textbf{Internal validity.}
% \outline{TODO: failure-predictor bias; whether branch isolation is truly leak-free (side-effect leakage).}

% \textbf{External validity.}
% \outline{TODO: representativeness of Multi-Docker-Eval; coverage of backbone models.}

% \textbf{Construct validity.}
% \outline{TODO: fairness of the success criterion and the wall-clock measurement protocol (hardware jitter, network variance).}

\label{sec:threats}

\noindent\textbf{Internal validity.} LLM-based agents can vary across model backends and runs, a common issue in recent software-engineering agent evaluations~\cite{11029914, xia2024agentless}.
We control for this by evaluating all methods with the same benchmark, budget, execution harness, and validation scripts within each comparison, and by checking that the main conclusion holds across three backbone models.
We also confirm the gains with paired significance tests (McNemar's test for Success Rate~\cite{mcnemar1947note}, Wilcoxon signed-rank for time~\cite{wilcoxon1945individual}; $p<0.005$).

\noindent\textbf{External validity.} We evaluate on Multi-Docker-Eval because it is designed for executable environment construction and covers 334 tasks from 40 repositories and 9 programming languages~\cite{fu-etal-2026-multi}.
This multilingual setting reduces the risk of tuning to one ecosystem, while broader industrial settings remain an important target for future evaluation.

\noindent\textbf{Construct validity.} Our two primary metrics match the paper's objectives: F2P measures whether the environment verifies the patch, and wall-clock time measures setup latency.
Following SWE-bench and Multi-Docker-Eval, F2P requires the tests to fail on the original code and pass after the gold patch~\cite{ jimenez2024swe,fu-etal-2026-multi}, the executable validation used by current SWE benchmarks and data pipelines~\cite{ pan2024training,jain2025r2e}.
We measure wall-clock time under identical hardware and matched concurrency, so infrastructure variation affects absolute times more than relative comparisons.

\section{Related Work}
\label{sec:related}

\subsection{Speculative Execution and Decoding}
Speculation is a classic latency-hiding principle in computer systems: perform work that may be needed before its necessity is known, commit it when the prediction is validated, and discard it otherwise.
This pattern appears in speculative computation and computer architecture~\cite{burton1985speculative, 10.1145/1411204.1411205}, thread-level speculation that executes likely future code in parallel~\cite{10.1145/2938369}, and optimistic concurrency control that validates tentative work before committing it~\cite{ 10.1145/319566.319567}.
\methodname{} follows the same systems principle, but applies it to software environment construction: the speculative unit is an isolated build branch, and the validation signal is real execution.

Speculation has recently become central to LLM inference through speculative decoding~\cite{leviathan2023fast, chen2023accelerating, li2026empirical}.
A lightweight draft model proposes future tokens, the target model verifies them in parallel, accepted tokens are committed, and rejected tokens are discarded.
Follow-up work improves this draft-and-verify loop with multiple decoding heads, feature-level drafting, and self-speculation~\cite{cai2024medusa, li2024eagle,zhang2024draft}.
More recent agent-serving work extends speculation beyond tokens, speculatively executing likely actions or tools to reduce the latency of sequential agent loops~\cite{ ye2025speculative,sui2026parallelizing, guan2025dynamic}.
Closest to our latency-hiding goal, EAGER overlaps LLM code generation with interpreter execution to hide execution latency during code interpretation~\cite{sun2026executing}.

\methodname{} instead speculates on setup routes and likely command failures, returning the first committed environment.
This makes speculation valuable in a new setting: long-running environment setup commands, where a correct prediction can remove an entire wait-diagnose-retry round from the critical path.

\subsection{Automated Environment Construction}
Environment setup has long been a software-engineering and DevOps problem rather than only a benchmark artifact.
Container specifications such as Dockerfiles are widely used and studied~\cite{cito2017empirical,henkel2020learning, eng2021revisiting}, but developers still report Docker and Docker-Compose specifications as time-consuming and knowledge-intensive~\cite{reis2021developing, rosa2023automatically}.
Prior work has therefore explored assistance for containerization, including dependency inference from source code~\cite{ 8811897}, Dockerfile generation from knowledge graphs~\cite{ 9529714}, Dockerfile completion~\cite{hanayama2020humpback}, base-image recommendation~\cite{9825856}, and Dockerfile generation for reproducible research~\cite{nust2019containerit}.
These works help produce or recommend environment specifications; \methodname{} focuses on the execution-time process of finding a specification that actually passes validation.

Executable environments are also central to modern software-engineering automation.
Benchmarks and training pipelines need tests that can run on real repositories~\cite{jimenez2024swe, pan2024training,jain2025r2e}, and recent environment benchmarks and datasets scale this need across many repositories and languages~\cite{fu-etal-2026-multi,eliseeva2025envbench, milliken2025beyond,chen2026swe,fu2026davinci}.
To automate this process, Repo2Run~\cite{hu2026repo2run} and ExecutionAgent~\cite{bouzenia2025you} iteratively issue commands and refine setup scripts; SWE-Builder~\cite{guo2025swe} automates environment construction inside an issue-resolution data pipeline; RepoLaunch~\cite{li2026repolaunch} resolves dependencies, builds repositories, and extracts test results across languages and platforms; and MEnvAgent~\cite{guo2026menvagent} reuses and patches retrieved environments instead of always rebuilding from scratch.

These systems share the execution-driven style common to LLM agents for software engineering~\cite{11029914,xia2024agentless,NEURIPS2024_5a7c9475, 10.1145/3650212.3680384}: run an action, observe feedback, and decide the next step.
\methodname{} keeps the same real-execution feedback, but changes the schedule: it explores alternative setup routes and likely failures while slow commands are still running.
This turns waiting time into useful search, targeting both fail-to-pass success and wall-clock time.

\section{Conclusion}
LLM-based environment setup remains slow and brittle because agents usually follow a sequential trial-and-error loop: they try one setup route, wait for slow commands to finish, and only then decide what to repair or try next.
This paper introduced \methodname{}, which treats fail-to-pass success and wall-clock time as joint objectives and uses speculative forking to turn waiting time into useful parallel search.
Speculative forking supports two granularities: strategy-level route exploration before execution and command-level repair preparation while the original command is still running.
On Multi-Docker-Eval, covering 334 tasks from 40 repositories and 9 languages, \methodname{} improves the Success Rate by 6.4 to 12.0 percentage points over the strongest baseline while reducing average setup time by 22.4\% to 41.6\% across three backbone models.
Speculative forking shows that LLM-based environment setup need not trade speed for reliability: exploring candidate branches in parallel and returning the first committed environment makes agents both faster and more successful.

\section*{Data Availability}
To support reproducibility, all related code, scripts, and experimental data are available at \url{https://anonymous.4open.science/r/Speculative-Forking-C907}.

% \outline{Optional section: Acknowledgment.}
% \outline{TODO: funding / acknowledgment.}

\bibliographystyle{IEEEtran}
\bibliography{references}

\begin{thebibliography}{10}
\providecommand{\url}[1]{#1}
\csname url@samestyle\endcsname
\providecommand{\newblock}{\relax}
\providecommand{\bibinfo}[2]{#2}
\providecommand{\BIBentrySTDinterwordspacing}{\spaceskip=0pt\relax}
\providecommand{\BIBentryALTinterwordstretchfactor}{4}
\providecommand{\BIBentryALTinterwordspacing}{\spaceskip=\fontdimen2\font plus
\BIBentryALTinterwordstretchfactor\fontdimen3\font minus
  \fontdimen4\font\relax}
\providecommand{\BIBforeignlanguage}[2]{{%
\expandafter\ifx\csname l@#1\endcsname\relax
\typeout{** WARNING: IEEEtran.bst: No hyphenation pattern has been}%
\typeout{** loaded for the language `#1'. Using the pattern for}%
\typeout{** the default language instead.}%
\else
\language=\csname l@#1\endcsname
\fi
#2}}
\providecommand{\BIBdecl}{\relax}
\BIBdecl

\bibitem{bouzenia2025you}
I.~Bouzenia and M.~Pradel, ``You name it, i run it: An llm agent to execute
  tests of arbitrary projects,'' \emph{Proceedings of the ACM on Software
  Engineering}, vol.~2, no. ISSTA, pp. 1054--1076, 2025.

\bibitem{henkel2020learning}
J.~Henkel, C.~Bird, S.~K. Lahiri, and T.~Reps, ``Learning from, understanding,
  and supporting devops artifacts for docker,'' in \emph{Proceedings of the
  ACM/IEEE 42nd international conference on software engineering}, 2020, pp.
  38--49.

\bibitem{cito2017empirical}
J.~Cito, G.~Schermann, J.~E. Wittern, P.~Leitner, S.~Zumberi, and H.~C. Gall,
  ``An empirical analysis of the docker container ecosystem on github,'' in
  \emph{2017 IEEE/ACM 14th International Conference on Mining Software
  Repositories (MSR)}.\hskip 1em plus 0.5em minus 0.4em\relax IEEE, 2017, pp.
  323--333.

\bibitem{eng2021revisiting}
K.~Eng and A.~Hindle, ``Revisiting dockerfiles in open source software over
  time,'' in \emph{2021 IEEE/ACM 18th International Conference on Mining
  Software Repositories (MSR)}.\hskip 1em plus 0.5em minus 0.4em\relax IEEE,
  2021, pp. 449--459.

\bibitem{reis2021developing}
D.~Reis, B.~Piedade, F.~F. Correia, J.~P. Dias, and A.~Aguiar, ``Developing
  docker and docker-compose specifications: A developers’ survey,''
  \emph{Ieee Access}, vol.~10, pp. 2318--2329, 2021.

\bibitem{rosa2023automatically}
G.~Rosa, A.~Mastropaolo, S.~Scalabrino, G.~Bavota, and R.~Oliveto,
  ``Automatically generating dockerfiles via deep learning: Challenges and
  promises,'' \emph{arXiv preprint arXiv:2303.15990}, 2023.

\bibitem{pan2024training}
J.~Pan, X.~Wang, G.~Neubig, N.~Jaitly, H.~Ji, A.~Suhr, and Y.~Zhang, ``Training
  software engineering agents and verifiers with swe-gym,'' \emph{arXiv
  preprint arXiv:2412.21139}, 2024.

\bibitem{yang2026swe}
J.~Yang, K.~Lieret, C.~Jimenez, A.~Wettig, K.~Khandpur, Y.~Zhang, B.~Hui,
  O.~Press, L.~Schmidt, and D.~Yang, ``Swe-smith: Scaling data for software
  engineering agents,'' \emph{Advances in Neural Information Processing
  Systems}, vol.~38, 2026.

\bibitem{jain2025r2e}
N.~Jain, J.~Singh, M.~Shetty, L.~Zheng, K.~Sen, and I.~Stoica, ``R2e-gym:
  Procedural environments and hybrid verifiers for scaling open-weights swe
  agents,'' \emph{arXiv preprint arXiv:2504.07164}, 2025.

\bibitem{yuan2026swe}
D.~Yuan, W.~Wu, Z.~Wang, X.~Zhao, H.~Zhang, and D.~Zhao, ``Swe-minisandbox:
  Container-free reinforcement learning for building software engineering
  agents,'' \emph{arXiv preprint arXiv:2602.11210}, 2026.

\bibitem{chen2026securevibebench}
J.~Chen, H.~Huang, Y.~Lyu, J.~An, J.~Shi, C.~Yang, T.~Zhang, H.~Tian, Y.~Li,
  Z.~Li \emph{et~al.}, ``Securevibebench: Benchmarking secure vibe coding of ai
  agents via reconstructing vulnerability-introducing scenarios,'' in
  \emph{Proceedings of the 64th Annual Meeting of the Association for
  Computational Linguistics (Volume 1: Long Papers)}, 2026, pp.
  24\,144--24\,168.

\bibitem{shi2026swe}
Y.~Shi, J.~Xu, K.~Fu, W.~Zeng, S.~He, L.~Zhang, Y.~Liu, Z.~Zhao, T.~Y. Zhuo,
  J.~Cao \emph{et~al.}, ``Swe-bench promax: Benchmarking agents on large-scale
  multilingual code refactoring,'' \emph{arXiv preprint arXiv:2608.09802},
  2026.

\bibitem{chen2026swe}
M.~Chen, L.~Zhang, Y.~Feng, X.~Wang, W.~Zhao, R.~Cao, J.~Yang, J.~Chen, M.~Li,
  Z.~Ma \emph{et~al.}, ``Swe-universe: Scale real-world verifiable environments
  to millions,'' \emph{arXiv preprint arXiv:2602.02361}, 2026.

\bibitem{fu2026davinci}
D.~Fu, S.~Wu, Y.~Wu, Z.~Peng, Y.~Huang, J.~Sun, J.~Zeng, M.~Jiang, L.~Zhang,
  Y.~Li \emph{et~al.}, ``davinci-env: Open swe environment synthesis at
  scale,'' \emph{arXiv preprint arXiv:2603.13023}, 2026.

\bibitem{hu2026repo2run}
R.~Hu, C.~Peng, J.~Xu, and C.~Gao, ``Repo2run: Automated building executable
  environment for code repository at scale,'' \emph{Advances in Neural
  Information Processing Systems}, vol.~38, pp. 32\,679--32\,718, 2026.

\bibitem{guo2025swe}
L.~Guo, Y.~Wang, C.~Li, W.~Tao, P.~Yang, J.~Chen, H.~Song, D.~Tang, and
  Z.~Zheng, ``Swe-factory: Your automated factory for issue resolution training
  data and evaluation benchmarks,'' \emph{arXiv preprint arXiv:2506.10954},
  2025.

\bibitem{li2026repolaunch}
K.~Li, R.~Li, L.~Zhang, Q.~Jin, L.~Zhu, X.~Huang, G.~Zhang, Y.~Zhang, S.~He,
  C.~Xie \emph{et~al.}, ``Repolaunch: Automating build\&test pipeline of code
  repositories on any language and any platform,'' \emph{arXiv e-prints}, pp.
  arXiv--2603, 2026.

\bibitem{guo2026menvagent}
C.~Guo, J.~Wu, S.~He, Y.~Chen, Z.~Kuang, S.~Fan, B.~Chen, S.~Bao, J.~Liu, H.~Wu
  \emph{et~al.}, ``Menvagent: Scalable polyglot environment construction for
  verifiable software engineering,'' \emph{arXiv preprint arXiv:2601.22859},
  2026.

\bibitem{11029914}
I.~Bouzenia, P.~Devanbu, and M.~Pradel, ``Repairagent: An autonomous, llm-based
  agent for program repair,'' in \emph{2025 IEEE/ACM 47th International
  Conference on Software Engineering (ICSE)}, 2025, pp. 2188--2200.

\bibitem{xia2024agentless}
C.~S. Xia, Y.~Deng, S.~Dunn, and L.~Zhang, ``Agentless: Demystifying llm-based
  software engineering agents,'' \emph{arXiv preprint arXiv:2407.01489}, 2024.

\bibitem{chen2026agentexecutor}
J.~Chen, C.~Yang, X.~Hu, Z.~Li, X.~Xia, and D.~Lo, ``Agentexecutor: Partial
  code execution via agentic context generation,'' \emph{arXiv preprint
  arXiv:2608.05959}, 2026.

\bibitem{fu-etal-2026-multi}
\BIBentryALTinterwordspacing
K.~Fu, T.~Liu, Z.~Shang, Y.~MA, J.~Liu, J.~Yang, and K.~Bian,
  ``Multi-docker-eval: A `shovel of the gold rush' benchmark on automatic
  environment building for software engineering,'' in \emph{Findings of the
  {A}ssociation for {C}omputational {L}inguistics: {ACL} 2026}, M.~Liakata,
  V.~P. Moreira, J.~Zhang, and D.~Jurgens, Eds.\hskip 1em plus 0.5em minus
  0.4em\relax San Diego, California, United States: Association for
  Computational Linguistics, Jul. 2026, pp. 17\,911--17\,927. [Online].
  Available: \url{https://aclanthology.org/2026.findings-acl.889/}
\BIBentrySTDinterwordspacing

\bibitem{deepseekai2026deepseekv4}
DeepSeek-AI, ``Deepseek-v4: Towards highly efficient million-token context
  intelligence,'' 2026.

\bibitem{mimo2026v25pro}
{Xiaomi MiMo Team}, ``Mimo-v2.5-pro,''
  \url{https://huggingface.co/collections/XiaomiMiMo/mimo-v25}, 2026.

\bibitem{openai2026gpt55}
{OpenAI}, ``{GPT-5.5 System Card},''
  \url{https://openai.com/index/gpt-5-5-system-card/}, 2026, accessed:
  2026-06-30.

\bibitem{mcnemar1947note}
Q.~McNemar, ``Note on the sampling error of the difference between correlated
  proportions or percentages,'' \emph{Psychometrika}, vol.~12, no.~2, pp.
  153--157, 1947.

\bibitem{wilcoxon1945individual}
F.~Wilcoxon, ``Individual comparisons by ranking methods,'' \emph{Biometrics
  bulletin}, vol.~1, no.~6, pp. 80--83, 1945.

\bibitem{10.1145/1411204.1411205}
\BIBentryALTinterwordspacing
B.~W. Lampson, ``Lazy and speculative execution in computer systems,'' in
  \emph{Proceedings of the 13th ACM SIGPLAN International Conference on
  Functional Programming}, ser. ICFP '08.\hskip 1em plus 0.5em minus
  0.4em\relax New York, NY, USA: Association for Computing Machinery, 2008, p.
  1–2. [Online]. Available: \url{https://doi.org/10.1145/1411204.1411205}
\BIBentrySTDinterwordspacing

\bibitem{leviathan2023fast}
Y.~Leviathan, M.~Kalman, and Y.~Matias, ``Fast inference from transformers via
  speculative decoding,'' in \emph{International Conference on Machine
  Learning}.\hskip 1em plus 0.5em minus 0.4em\relax PMLR, 2023, pp.
  19\,274--19\,286.

\bibitem{ye2025speculative}
N.~Ye, A.~Ahuja, G.~Liargkovas, Y.~Lu, K.~Kaffes, and T.~Peng, ``Speculative
  actions: A lossless framework for faster agentic systems,'' \emph{arXiv
  preprint arXiv:2510.04371}, 2025.

\bibitem{jimenez2024swe}
C.~E. Jimenez, J.~Yang, A.~Wettig, S.~Yao, K.~Pei, O.~Press, and K.~Narasimhan,
  ``Swe-bench: Can language models resolve real-world github issues?'' in
  \emph{International Conference on Learning Representations}, vol. 2024, 2024,
  pp. 54\,107--54\,157.

\bibitem{burton1985speculative}
F.~W. Burton, ``Speculative computation, parallelism, and functional
  programming,'' \emph{IEEE Transactions on Computers}, vol. 100, no.~12, pp.
  1190--1193, 1985.

\bibitem{10.1145/2938369}
\BIBentryALTinterwordspacing
A.~Estebanez, D.~R. Llanos, and A.~Gonzalez-Escribano, ``A survey on
  thread-level speculation techniques,'' \emph{ACM Comput. Surv.}, vol.~49,
  no.~2, Jun. 2016. [Online]. Available: \url{https://doi.org/10.1145/2938369}
\BIBentrySTDinterwordspacing

\bibitem{10.1145/319566.319567}
\BIBentryALTinterwordspacing
H.~T. Kung and J.~T. Robinson, ``On optimistic methods for concurrency
  control,'' \emph{ACM Trans. Database Syst.}, vol.~6, no.~2, p. 213–226,
  Jun. 1981. [Online]. Available: \url{https://doi.org/10.1145/319566.319567}
\BIBentrySTDinterwordspacing

\bibitem{chen2023accelerating}
C.~Chen, S.~Borgeaud, G.~Irving, J.-B. Lespiau, L.~Sifre, and J.~Jumper,
  ``Accelerating large language model decoding with speculative sampling,''
  \emph{arXiv preprint arXiv:2302.01318}, 2023.

\bibitem{li2026empirical}
Y.~Li, J.~Chen, X.~Hu, and X.~Xia, ``An empirical study of speculative decoding
  on software engineering tasks,'' \emph{arXiv preprint arXiv:2604.26469},
  2026.

\bibitem{cai2024medusa}
T.~Cai, Y.~Li, Z.~Geng, H.~Peng, J.~D. Lee, D.~Chen, and T.~Dao, ``Medusa:
  Simple llm inference acceleration framework with multiple decoding heads,''
  \emph{arXiv preprint arXiv:2401.10774}, 2024.

\bibitem{li2024eagle}
Y.~Li, F.~Wei, C.~Zhang, and H.~Zhang, ``Eagle: Speculative sampling requires
  rethinking feature uncertainty,'' \emph{arXiv preprint arXiv:2401.15077},
  2024.

\bibitem{zhang2024draft}
J.~Zhang, J.~Wang, H.~Li, L.~Shou, K.~Chen, G.~Chen, and S.~Mehrotra, ``Draft\&
  verify: Lossless large language model acceleration via self-speculative
  decoding,'' in \emph{Proceedings of the 62nd Annual Meeting of the
  Association for Computational Linguistics (Volume 1: Long Papers)}, 2024, pp.
  11\,263--11\,282.

\bibitem{sui2026parallelizing}
Y.~Sui, H.~Zhao, R.~Ma, Z.~He, H.~Wang, J.~Li, K.~Xu, K.~Chen, and Y.~Yang,
  ``Parallelizing tool execution and llm generation for low-latency agent
  serving,'' \emph{arXiv preprint arXiv:2603.18897}, 2026.

\bibitem{guan2025dynamic}
Y.~Guan, Q.~Lan, S.~Fei, D.~Ding, D.~Acharya, C.~Wang, W.~Y. Wang, and W.~Hua,
  ``Dynamic speculative agent planning,'' \emph{arXiv preprint
  arXiv:2509.01920}, 2025.

\bibitem{sun2026executing}
Z.~Sun, Z.~Lin, Z.~Chen, C.~Yang, M.~Zhou, L.~Li, and D.~Lo, ``Executing as you
  generate: Hiding execution latency in llm code generation,'' \emph{arXiv
  preprint arXiv:2604.00491}, 2026.

\bibitem{8811897}
E.~Horton and C.~Parnin, ``Dockerizeme: Automatic inference of environment
  dependencies for python code snippets,'' in \emph{2019 IEEE/ACM 41st
  International Conference on Software Engineering (ICSE)}, 2019, pp. 328--338.

\bibitem{9529714}
H.~Ye, J.~Zhou, W.~Chen, J.~Zhu, G.~Wu, and J.~Wei, ``Dockergen: A knowledge
  graph based approach for software containerization,'' in \emph{2021 IEEE 45th
  Annual Computers, Software, and Applications Conference (COMPSAC)}, 2021, pp.
  986--991.

\bibitem{hanayama2020humpback}
K.~Hanayama, S.~Matsumoto, and S.~Kusumoto, ``Humpback: Code completion system
  for dockerfiles based on language models,'' in \emph{Proc. Workshop on
  Natural Language Processing Advancements for Software Engineering}, vol.
  2020, 2020, pp. 1--4.

\bibitem{9825856}
Y.~Zhang, Y.~Zhang, X.~Mao, Y.~Wu, B.~Lin, and S.~Wang, ``Recommending base
  image for docker containers based on deep configuration comprehension,'' in
  \emph{2022 IEEE International Conference on Software Analysis, Evolution and
  Reengineering (SANER)}, 2022, pp. 449--453.

\bibitem{nust2019containerit}
D.~N{\"u}st and M.~Hinz, ``containerit: Generating dockerfiles for reproducible
  research with r,'' \emph{Journal of Open Source Software}, vol.~4, no.~40, p.
  1603, 2019.

\bibitem{eliseeva2025envbench}
A.~Eliseeva, A.~Kovrigin, I.~Kholkin, E.~Bogomolov, and Y.~Zharov, ``Envbench:
  A benchmark for automated environment setup,'' \emph{arXiv preprint
  arXiv:2503.14443}, 2025.

\bibitem{milliken2025beyond}
L.~Milliken, S.~Kang, and S.~Yoo, ``Beyond pip install: Evaluating llm agents
  for the automated installation of python projects,'' in \emph{2025 IEEE
  International Conference on Software Analysis, Evolution and Reengineering
  (SANER)}.\hskip 1em plus 0.5em minus 0.4em\relax IEEE, 2025, pp. 1--11.

\bibitem{NEURIPS2024_5a7c9475}
\BIBentryALTinterwordspacing
J.~Yang, C.~Jimenez, A.~Wettig, K.~Lieret, S.~Yao, K.~Narasimhan, and O.~Press,
  ``Swe-agent: Agent-computer interfaces enable automated software
  engineering,'' in \emph{Advances in Neural Information Processing Systems},
  A.~Globerson, L.~Mackey, D.~Belgrave, A.~Fan, U.~Paquet, J.~Tomczak, and
  C.~Zhang, Eds., vol.~37.\hskip 1em plus 0.5em minus 0.4em\relax Curran
  Associates, Inc., 2024, pp. 50\,528--50\,652. [Online]. Available:
  \url{https://proceedings.neurips.cc/paper_files/paper/2024/file/5a7c947568c1b1328ccc5230172e1e7c-Paper-Conference.pdf}
\BIBentrySTDinterwordspacing

\bibitem{10.1145/3650212.3680384}
\BIBentryALTinterwordspacing
Y.~Zhang, H.~Ruan, Z.~Fan, and A.~Roychoudhury, ``Autocoderover: Autonomous
  program improvement,'' in \emph{Proceedings of the 33rd ACM SIGSOFT
  International Symposium on Software Testing and Analysis}, ser. ISSTA
  2024.\hskip 1em plus 0.5em minus 0.4em\relax New York, NY, USA: Association
  for Computing Machinery, 2024, p. 1592–1604. [Online]. Available:
  \url{https://doi.org/10.1145/3650212.3680384}
\BIBentrySTDinterwordspacing

\end{thebibliography}

\end{document}